\documentclass{aa}
\usepackage{graphicx}
\usepackage{txfonts}

\newcommand{\eV}{{\rm e\hspace{-1pt}V}}
\newcommand{\DHIP}{$\Delta_{\rm HIP}$}
\newcommand{\FeI}{\ion{Fe}{i}}
\newcommand{\FeII}{\ion{Fe}{ii}}
\newcommand{\MgI}{\ion{Mg}{i}}
\newcommand{\AlI}{\ion{Al}{i}}
\newcommand{\HIP}{{\sc Hipparcos}}
\newcommand{\logg}{$\log g$}
\newcommand{\m}{$-$}
\newcommand{\SH}{S$_{\rm H}$}
\newcommand{\Teff}{$T_{\rm eff}$}

\begin{document}

\title{Kinetic equilibrium of iron in the atmospheres of cool stars}
\subtitle{III. The ionization equilibrium of selected reference stars
\thanks{Based on observations collected at the German-Spanish Astro\-nomical
Centre, Calar Alto, Spain}}
\titlerunning{Kinetic equilibrium of iron in the atmospheres of reference stars}
\author{Andreas J. Korn\inst{1,2}\and Jianrong Shi\inst{1,3}
\and Thomas Gehren\inst{1}}
\offprints{A.J. Korn,\\ \email{ajkorn@usm.uni-muenchen.de}}
\institute{Institut f{\"u}r Astronomie und Astrophysik der Universit\"at
M{\"u}nchen,\\ Universit{\"a}ts-Sternwarte M{\"u}nchen (USM), Scheinerstra{\ss}e 1, D-81679
M{\"u}nchen, Germany \and
Max-Planck Institut f{\"u}r extraterrestrische Physik (MPE), Giessenbachstra{\ss}e, D-85748 Garching \and
National Astronomical Observatories, Chinese Academy of Sciences (NAOC), Beijing 100012, China}
\date{Received / Accepted}
\abstract{%
Non-LTE line formation calculations of \FeI\ are performed for a small number of reference stars to investigate and quantify the efficiency of neutral hydrogen collisions. Using the atomic model that was described in previous publications, the final discrimination with respect to hydrogen collisions is based on the condition that the surface gravities as determined by the \FeI/\FeII\ ionization equilibria are in agreement with their astrometric counterparts obtained from \HIP\ parallaxes.\\
High signal-to-noise, high-resolution \'{e}chelle spectra are analysed to determine individual profile fits and differential abundances of iron lines. Depending on the choice of the hydrogen collision scaling factor \SH, we find deviations from LTE in \FeI\ ranging from 0.00 (\SH\,=\,$\infty$) to 0.46\,dex (\SH\,=\,0 for HD140283) in the logarithmic abundances while \FeII\ follows LTE. With the exception of Procyon, for which a mild temperature correction is needed to fulfil the ionization balance, excellent consistency is obtained for the metal-poor reference stars if Balmer profile temperatures are combined with \SH\,=\,3. This value is much higher than what is found for simple atoms like Li or Ca, both from laboratory measurements and inference of stellar analyses.\\
The correct choice of collisional damping parameters (''van-der-Waals'' constants) is found to be generally more important for these little evolved metal-poor stars than considering departures from LTE. For the Sun the calibrated value for \SH\ leads to average \FeI\ non-LTE corrections of 0.02\,dex and a mean abundance from \FeI\ lines of $\log$\,$\varepsilon$(Fe)\,=\,7.49\,$\pm$\,0.08.\\
We confront the deduced stellar parameters with comparable spectroscopic analyses by other authors which also rely on the iron ionization equilibrium as a gravity indicator. On the basis of the \HIP\ astrometry our results are shown to be an order of magnitude more precise than published data sets, both in terms of offset and star-to-star scatter.
 }
\maketitle

\section{Introduction}
As is evident from our previous investigations (Gehren et al. \cite{GBMRS01},
\cite{GKS02}), the large scatter among laboratory \emph{f} values of neutral iron lines
does not allow the unambiguous determination of a kinetic model of the iron atom
in the solar atmosphere. In particular, the r\^{o}le of collisions with neutral
hydrogen atoms could not  be fully explored. This resulted in an uncertainty of
the hydrogen collision scaling factor \SH\ (a multiplicative modification to the Drawin (\cite{Drawin68}, \cite{Drawin69}) formula for allowed inelastic collisions with neutral hydrogen) which becomes important as soon as the kinetic model is applied to metal-poor stars. Since in the atmospheres of halo
and thick disk stars line-blanketing in the UV becomes more and more unimportant
leading to an increase of photoionization, while simultaneously electron
collisions can no longer counterbalance the predominance of radiation interactions,
the hydrogen collisions are an essential part of the kinetic equilibrium of
iron, and many other photoionization-dominated metals as well. This problem is further enhanced by the new radiative cross-sections of Bautista (\cite{Bautista97}) which are
substantially larger than simple hydrogenic approximations would make believe.

To establish the influence of deviations from LTE on the \emph{ionization
equilibrium} of iron we have therefore chosen a number of {\em reference stars}
which we feel are sufficiently representative of the local metal-poor
stellar generation on and near the main sequence. Our interest is particularly devoted to the
important question: to what extent is the derivation of spectroscopic stellar parameters and metallicities affected by the kinetic equilibrium of iron? Answering this question involves the investigation of the radiation field (depending mostly on the effective temperature) and of the collisional
interaction (depending mostly on surface gravity). Therefore our choice of
metal-poor reference stars has led to four representatives: HD~103095 ($\equiv$ Gmb~1830) as a cool
main sequence star, HD~19445 as a typical subdwarf, HD 84937 as a turnoff star and HD~140283 as a moderately cool subgiant. The last two of these
reference stars have also been chosen by Asplund et al. (\cite{ANTS00}) to model
hydrodynamic convection in metal-poor stars.
\begin{table}
\caption[]{Basic stellar parameters of the reference stars. The iron abundance is derived
from \FeII\ lines with the microturbulence fulfilling the usual line strength constraint (see text). [$\alpha$/Fe] = 0.4 was adopted in the computation of the metal-poor atmospheres, for Procyon ($\equiv$ HD 61421) a solar mixture was assumed. Except for Procyon (data from Allende Prieto et al. \cite{AAGL02}), surface gravities are from \HIP\ parallaxes with masses derived from the tracks of VandenBerg et al. (\cite{VSRIA00}). [Fe/H]\,=\,$\log($Fe/H$)_\star - \log($Fe/H$)_\odot$}
\label{params}
\begin{tabular}{lrrrrrr}
 & \Teff &$\log g$&[Fe/H]&$\xi$& $\pi_{\rm HIP}$ & $M$\\
star & [K] & (cgs) & &[km/s]& [\arcsec] & [M$_\odot$] \\
\noalign{\smallskip}\hline \noalign{\smallskip}
HD 103095  & 5070  & 4.69  & \m1.35 & 0.95 & 109.2 & 0.62 \\
HD 19445  & 6032  & 4.40  & \m2.08 & 1.75 & 25.9 & 0.67 \\
HD 140283$^1$ & 5806  & 3.69  & \m2.42 & 1.70 & 17.4 & 0.79 \\
HD 84937$^1$  & 6346  & 4.00  & \m2.16 & 1.80 & 12.4 & 0.79 \\
Procyon   & 6510  & 3.96  & \m0.03 & 1.83 & 285.9 & 1.42 \\
\noalign{\smallskip}\hline\noalign{\smallskip}
Sun & 5780 & 4.44 & 0.0 & 1.0 & --- & 1.00 \\
\noalign{\smallskip}\hline\noalign{\smallskip}
\end{tabular}
$^1$ for this star extinction due to interstellar reddening was considered following Hauck \& Mermilliod (\cite{HM98}, cf. Table~\ref{stellarparameters})
\end{table}

Photospheric surface gravities are usually determined by means of three independent methods. In
principle the most reliable approach uses {\rm astrometry} (presently the {\sc Hipparcos} parallaxes, ESA \cite{ESA97}) with an appropriate guess of the stellar mass, which enters the resulting $\log g$ only
with a $\sigma(\log M)$ uncertainty. Since the mass uncertainty of these very old stars is
generally below 0.05\,M$_{\odot}$ (when referring to a given set of evolutionary tracks), the corresponding error stays below 0.03.
The uncertainty of the effective temperature enters as $4\,\sigma$(log \Teff)$\,\approx\,0.02$.
Thus the prime error in $\log g$ is due to that of the parallax itself ($2\,\sigma(\log \pi)$),
which for the four stars is between 0.01 (Gmb~1830) and 0.07 (HD~84937). The maximum error then is
between 0.06 and 0.12 in $\log g$.

The second method refers to the {\em damping wings} of strong neutral metal
lines such as the \ion{Mg}{i}b triplet. However, as demonstrated by Fuhrmann
(\cite{Fuhrmann98a}), for stars with [Fe/H]\,$<$\,$-2$ the wings of such lines become
extremely shallow and the determination correspondingly uncertain. The combination of Balmer profile temperatures and Mg\,{\sc i}b gravities (both in LTE) was shown to be fully compatible with the \HIP\ astrometry (Fuhrmann \cite{Fuhrmann98}, \cite{Fuhrmann03}).

Originally, the surface gravity of remote subdwarfs and subgiants was derived from the
{\em ionization equilibrium} of iron where for simplicity and lack of better evidence one used
the assumption that \FeI/\FeII\ is populated according to the Saha/Boltzmann
equilibrium (LTE). For a number of stars including Procyon as a primary standard this
was shown to lead to results incompatible with those obtained from strong line
wings (Fuhrmann et al. \cite{FPFRG97}) or \HIP\ parallaxes. More specifically, evolutionary stages are predicted which have not yet been reached. Therefore,
the present approach has abandoned the LTE assumption and obtains the surface
gravity from a fit to the Fe lines in kinetic equilibrium.

To evaluate the reliability of the non-LTE calculations in stars with different
parameters Procyon ($\equiv \alpha$ CMi) was also selected, because the discrepancy
between $\log g$ obtained from ionization equilibrium and parallax
remains unresolved.

Table \ref{params} lists the basic stellar parameters of the five stars plus the Sun.

\section{Observations}

Spectra of the stars were taken with the fibre-fed \'echelle spectrograph FOCES (Pfeiffer et al. \cite{PFBFG98}) at the 2.2m telescope of the German-Spanish Astronomical Centre on Calar Alto during a number of observing runs between 1999 and 2001, with a spectral resolution of $R$\,=\,65\,000 (HD~140283 was observed at $R$\,=\,40\,000). As measured in line-free spectral windows near H$\alpha$, the peak signal-to-noise ratio (S/N) exceeds 300:1 in all cases. With a spectral coverage of 4200 -- 9000 \AA\ our analyses naturally focus on the lines of \FeI\ and \FeII\ in the visible disregarding the resonance lines in the (near) UV.\\
Owing to the design of FOCES (the light of both star and flatfield lamp being transmitted through the same optical path), the resulting flatfield calibration of the stellar spectra produces an extremely
predictable continuum run that contains only the remaining second-order variation with wavelength which results from the difference between stellar \Teff\ and the
temperature of the halogen lamp. The internal accuracy of the effective
temperature determination is therefore very high, with profile residuals well
below 0.5\,\% of the continuum (cf. Korn \cite{Korn02}). The profile analysis of metal lines also profits from this quality, especially when the line density is high.

\begin{table}
\caption[]{Comparison of effective temperature determinations for the five reference stars. TI99 = Th\'evenin \& Idiart (\cite{TI99}), F00 = Fulbright (\cite{Fulbright00}), AAM-R96 = Alonso et al. (\cite{AAMR96}), GCC96 = Gratton et al. (\cite{GCC96}), KSG03 = this work. The mean offset is computed for the four metal-poor stars according to $\Delta$\,\Teff(study\,$-$\,KSG03). Significant offsets with respect to our temperature scale are encountered in the studies of Th\'evenin \& Idiart (\cite{TI99}) and Fulbright (\cite{Fulbright00})}
\label{teffs}
\begin{tabular}{lccccc}
star & TI99 & F00 &$\!\!$AAM-R96$\!\!$&GCC96&KSG03\\
 & $B\!-\!V$ & EE$^1$ & IRFM & IRFM &Balmer\\
\noalign{\smallskip}\hline \noalign{\smallskip}
HD 103095  & 4990 & 4950 & 5029 & 5124 \ \,& 5070 \\
HD 19445  & 5860 & 5825 & 6050 & 6066$^2$ & 6032 \\
HD 140283 & 5600 & 5650 & 5691 & 5766$^3$ & 5806 \\
HD 84937  & 6222 & 6375 & 6330 & 6351$^2$ & 6346 \\
mean offset & \m146\,\, & \m114\,\, & \m39 & +13 & $\pm$0$\!\!\!\!$ \\
\noalign{\smallskip}\hline \noalign{\smallskip}
Procyon   & 6631 &  --- & 6579 &  --- & 6508 \\
\noalign{\smallskip}\hline\noalign{\smallskip}
\end{tabular}
$^1$ EE = excitation equilibrium of \FeI\ in LTE\\
$^2$ average \Teff\ of two analyses\\
$^3$ average \Teff\ of three analyses
\end{table}

\section{Stellar effective temperatures and gravities}
The temperature scale adopted is based on Balmer line-broadening theory combined
from Stark effect profiles of Vidal et al. (\cite{VCS70}) and the Ali \& Griem
(\cite{AG65}, \cite{AG66}) approximation of resonance broadening (see Fuhrmann et
al. \cite{FAG93}). The basis of this approach was the different relative
contributions of these broadening processes to H$\alpha$ on the one hand
(essentially broadened by H collisions) and the higher series members (dominated
by the Stark effect) on the other. The results have been fixed with a minimum number of free
parameters for solar-type stars with ranging metallicities based on empirical evidence. Weights are assigned to the temperatures derived from each line based on temperature sensitivity, S/N and blaze correction reliability.
At present we refrain from adopting the recent hydrogen line-broadening theory of
Barklem et al. (\cite{BPO00}). As our own implementation shows, it does not reproduce the solar
temperature to within 100\,K (using the MARCS code and a $\chi^2$ fitting technique which allows for the intersection of observed profiles this value is 50\,K, Barklem et al. (\cite{BPO00}). In addition, it fails to reproduce the profile shape of higher Balmer lines (H$\beta$ and up) of stars like HD~140283 and HD~19445, both of which are located in the temperature band where differences between the two broadening theories are expected to be largest.


We utilize the atmospheric model MAFAGS (Fuhrmann et al. \cite{FPFRG97}) which is line-blanketed by means of the Kurucz ODFs (Kurucz \cite{K92}). Note that these ODFs were computed assuming\footnote{$\log$\,$\varepsilon$(Fe)\,=\,$\log\,(n_{\rm Fe}/n_{\rm H})\,+\,12$} $\log\,\varepsilon$(Fe)\,=\,7.67. We therefore rescale them by \m0.16\,dex to account for the low Solar iron abundance of $\log\,\varepsilon$(Fe)\,=\,7.51. Unlike in other widely-used atmospheric codes like ATLAS or MARCS, we use a mixing length $\alpha$\,=\,$l/H_{\rm p}$\,=\,0.5 in the framework of the B\"ohm-Vitense (\cite{BV58}) theory of convection in order to bring H$\alpha$ and the higher Balmer lines into concordance. The differences with respect to other methods of temperature determination have been described in Fuhrmann et al. (\cite{FAG93}) and Fuhrmann (\cite{Fuhrmann98}).
We note that the choice of the mixing-length affects the Balmer lines, but the influence on the \FeI\ and \FeII\ lines analysed in the Sun turns out to be equivalent to $\Delta\log\varepsilon_{{\rm Fe}} < 0.01$\,dex. The same is true for metal-poor stellar atmospheres such as that of HD\,140283. It also holds for the corresponding non-LTE computations.

The temperature scale resulting from our hydrogen line profile fitting is {\em higher} for the metal-poor stars than the photometric one of Th\'evenin \& Idiart (\cite{TI99}) and Fulbright (\cite{Fulbright00}) by more than 100\,K. It is, however, in excellent agreement with the infrared flux method (IRFM) temperatures of Alonso et al. (\cite{AAMR96}) and Gratton et al. (\cite{GCC96}). Table 2 confronts the corresponding numbers in detail. Only HD~140283 displays a markedly lower temperature in the Alonso et al. calibration. This is, at least in part, due to the zero-reddening assumption adopted by these authors. A discussion of this point can be found in Fuhrmann (\cite{Fuhrmann98a}).

Yet another independent temperature scale seems to support the Balmer profile temperatures: Peterson et al. (\cite{PDR01}) derive temperatures for HD~19445 and HD~84937 (and other stars) from absolute fluxes in the mid-UV. Effective temperatures of 6050\,K and 6300\,K were obtained. Peterson (2003, priv. comm.) analysed HD~140283 with similar methods obtaining 5850\,K. In total the mean offset between our temperature scales is thus well below 10\,K.

For the calibration the surface gravity is taken from {\sc Hipparcos} parallaxes. Additional quantities entering the equation are taken from Alonso et al. (\cite{AAMR96}, bolometric correction ${\rm BC}$), Hauck \& Mermilliod (\cite{HM98}, interstellar extinction $A_V$) and  VandenBerg et al. (\cite{VSRIA00}, stellar mass $M$). To derive the latter, the star is placed among the evolutionary tracks appropriately chosen with respect to metallicity ([Fe/H]$_{\rm NLTE}$) and $\alpha$-enhancement ([O/Fe]$_{\rm NLTE}$, according to Reetz \cite{Reetz99}). This process is iterated to convergence.

After the calibration the final atomic model (which best reproduces the astrometric results {\em on average}) is used to derive gravities which can then be compared with {\sc Hipparcos} to quantify the star-to-star scatter.

\section{Metal-poor reference stars and Procyon}
As described in Papers I and II in considerable detail, our present approach to
the analysis of the reference stars is based on a number of assumptions.

Atmospheric models are calculated for all stars (including the Sun) according to
the same type of input parameters, boundary conditions, and constraint
equations. We thus intentionally avoid the use of empirical models.

The major orientation of our analysis is that of \emph{empirical} evidence.
Although some theoretical model descriptions appear more satisfactory than
others, we build upon the experience that shortcomings in theories can be
detected only by comparison with observations.

Our basic approach to stellar spectroscopy is differential in the sense that we
require all types of input physics to be in agreement with solar observations.
This requirement clearly dictates the individual line fit approach, and it excludes the use of absolute $f$ values. It is important to recognize that the typical abundance scatter obtained from solar
spectral analysis of Fe lines is considerably higher than warranted by both
observational error and differences in theoretical models. There are essentially
two explanations for this result: either the $f$ values are not reliable or a
significant fraction of Fe lines are contaminated by relatively strong (yet undetected) blends.

Differential spectrum analysis must account for line broadening, particularly
when solar-type and extremely metal-poor stars are to be compared. This is in
fact the most important constraint because it involves additional atomic
parameters and line broadening theories. Every strategy to deal with strong line
wings automatically involves the problems of atmospheric modelling. Consequently,
for interpretation we establish steps of decreasing reliability with constraints coming from the observed line profile and the concept that the solar photospheric abundance of Fe is the same for all lines. Purely radiative parameters ($f$ values) are complemented by collisional broadening (Stark effect, van der
Waals-type damping). The variety of line strengths observed in the solar
spectrum then provides a comparison between weak and strong lines of the
resulting abundances, with all $f$ values taken from laboratory measurements and
the damping constants adjusted for the strong lines to minimize trends with line strength (see Paper II). Although this interferes occasionally with thermal or microturbulent broadening
velocities, Doppler broadening can be generally separated from the influence of
damping wings. As a result, stellar line abundances are compared with the
individual abundances of their solar counterparts. This is what is often called a line-by-line differential (non-)LTE analysis.

The line broadening problems are less significant for a differential analysis of
Fe II lines, however, here Doppler broadening plays an important role. The
scatter introduced to the solar \FeII\ abundance by the various data sets
is even higher than for \FeI\, thus a differential approach is absolutely
necessary.

The above considerations of the different contributions to line formation
severely limit the possible conclusions. Still, experience with stellar analyses show that atomic data are in many cases better determined from the solar spectrum than from a terrestrial laboratory.

\subsection{Non-LTE calculations}
With little progress in the understanding of the physics of hydrogen collisions, the review of Lambert (\cite{Lambert93}) still reflects our current state of knowledge. While the energy range encountered in stellar atmospheres is not accessible to experiment, Lambert concludes from measurements at 15\,\eV\ that the Drawin (\cite{Drawin68}, \cite{Drawin69}) formula most likely overestimates the true efficiency of this thermalizing process by two orders of magnitude. If this was the case, they could safely be disregarded. Rather than assuming that hydrogen collisions are inefficient all together, we take on the view of Steenbock \& Holweger (\cite{SH84}) that the "{\sl lack of reliable collision cross-sections constitutes a major problem in statistical-equilibrium studies of cool stars}". To face this problem we keep the efficiency of hydrogen collisions \SH\ as a free parameter which must be determined by appropriate observations.

As would be predicted, electron collisions cannot contribute very
much to thermalization, since the density of free electrons is roughly a factor
10 smaller than in comparable stars of solar metal content. Our calculations support this idea fully.
In metal-poor stars therefore no other process but hydrogen collisions can -- at least to some
extent and depending on their cross-section -- compensate for the high photoionization rates which would otherwise lead to extreme overionization at variance with the spectroscopic evidence given below. It is of
course unsatisfactory to use such a rather primitive representation of the collision process as the Drawin approximation which was never meant to be extended to anything else but collisions between rare gases. There is, however, a similar behaviour of both formulae of van Regemorter (\cite{VR62}, for
electron collisions) and Drawin (for hydrogen collisions) in that the rates for
both types of thermalization are proportional to the $f$ value and both depend
on the inverse threshold energy. Such a variation leads to a general increase of
collision rates with excitation energy, because highly excited levels cluster
near the ionization limit. Solar photospheric emission lines show at least for
\ion{Mg}{i} and \ion{Al}{i} that it is just these highly excited levels which
lead to Rydberg transitions dominated by a population inversion. This requires
low collision rates (whether electron or hydrogen) for such transitions, and in
fact the corresponding hydrogen collision factors for such atomic models were
found to fit the observed emission lines with \SH\ values significantly below
unity (Baum\"uller \& Gehren \cite{BG96}, Zhao et al. \cite{ZBG98}). Carlsson et al. (\cite{CRS92}) reproduce the IR emission lines entirely without hydrogen collisions, but do not simultaneously model the optical absorption lines.

\begin{table*}
\caption{\FeI\ and \FeII\ abundances for the reference stars and the Sun for different assumptions concerning the efficiency of hydrogen collisions. Differential abundances [Fe/H]$_{\rm diff}$ are given, except in the case of the Sun where log\,$\varepsilon$(Fe) is specified. $\sigma$ refers to the line-to-line scatter which is a factor $\sqrt n$ larger than the error of the mean. Note that the scatter in [\FeII/H]$_{\rm diff}$ is significantly reduced. Like in the case of HD~140283 discussed above, a strong imbalance between \FeI\ and \FeII\ results for all metal-poor stars if \SH\,=\,0 is assumed. Instead, the ionization equilibrium points towards \SH\,=\,3. Procyon cannot be fitted by any model} \label{sh0sh3lte}
\begin{center}
\begin{tabular}{rrrrcrc}  object     &  \SH\,=\,0  & \SH\,=\,3 & LTE & \# of & LTE & \# of \\
 & [\FeI/H]\,$\pm$\,$\sigma$ & [\FeI/H]\,$\pm$\,$\sigma$ & [\FeI/H]\,$\pm$\,$\sigma$ & lines & [\FeII/H]\,$\pm$\,$\sigma$ & lines \\[1mm]
\noalign{\smallskip}\hline \noalign{\smallskip}\\[-3mm]
HD 103095 & \m1.21\,$\pm$\,0.03 & \m1.35\,$\pm$\,0.04 & \m1.33\,$\pm$\,0.04 & 41 & \m1.36\,$\pm$\,0.03 & 15 \\[1mm]
HD19445 & \m1.67\,$\pm$\,0.07 & \m2.07\,$\pm$\,0.08 & \m2.07\,$\pm$\,0.08 & 58 & \m2.08\,$\pm$\,0.05 & 16 \\[1mm]
HD 140283 & \m2.00\,$\pm$\,0.07 & \m2.43\,$\pm$\,0.09 & \m2.46\,$\pm$\,0.09 & 62 & \m2.43\,$\pm$\,0.05 & 13 \\[1mm]
HD 84937 & \m1.78\,$\pm$\,0.07 & \m2.15\,$\pm$\,0.09 & \m2.18\,$\pm$\,0.09 & 61 & \m2.16\,$\pm$\,0.05 & 16 \\[1mm]
\hline\\[-2mm]
Procyon & \m0.08\,$\pm$\,0.06 & \m0.10\,$\pm$\,0.08 & \m0.11\,$\pm$\,0.07 & 52 & \m0.03\,$\pm$\,0.04 & 23 \\[1mm]
Sun & 7.54\,$\pm$\,0.08 & 7.49\,$\pm$\,0.08 & 7.46\,$\pm$\,0.07 & 96 & 7.53\,$\pm$\,0.10 & 37 \\
\noalign{\smallskip}\hline \noalign{\smallskip}
\end{tabular}
\end{center}
\end{table*}

\begin{figure}
\begin{center}
\vspace*{.5cm}
\includegraphics[angle=90,width=0.47\textwidth,clip=]{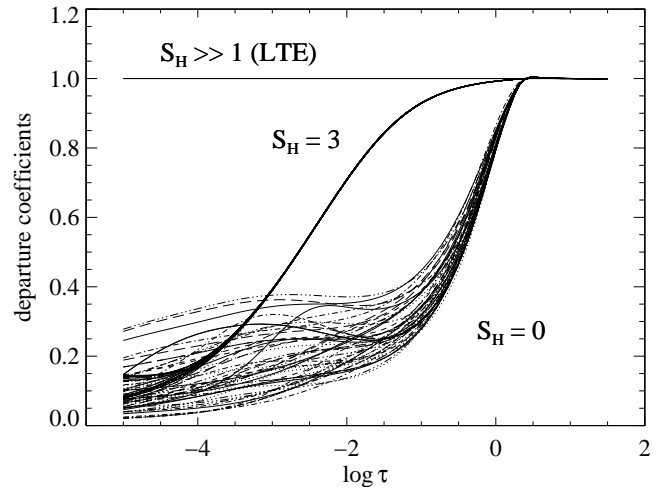}
\caption{Departure coefficients $b_i$ for \FeI\ terms in HD~140283 as a function of the optical depth $\log\,\tau_{5000}$. It becomes evident that they depend drastically on \SH. The line source functions are Planckian for all models with \SH\,$>$\,0, but not for the \SH\,=\,0 model. The best model (\SH\,=\,3) already returns abundances quite close to LTE}
\label{depcoeffs140283}
\end{center}
\end{figure}

For Fe, we determine \SH\ by requiring that both \FeI\ and \FeII\ lines produce a single-valued mean iron abundance when the gravity is assumed to be that derived from the \HIP\ parallax. This leaves the hydrogen collision enhancement factor \SH\ as the single free parameter in this investigation.

A number of non-LTE test calculations with various choices for \SH\ was performed for the reference stars, together with LTE calculations. Line profiles were fitted using external Gaussian broadening profiles essentially defined by the instrumental resolution. While the intrinsic profiles are better reproduced by a radial-tangential approximation to macroturbulence (Gray \cite{Gray77}), the dominance of the instrumental resolution and the relative weakness of the iron lines considered ($W_\lambda$\,$<$\,100\,m\AA) justify the use of a Gaussian. Equivalent widths (EWs) are then computed from the theoretical spectra to constrain the microturbulence $\xi$. Thus EWs only enter the analysis for the derivation of $\xi$, all other aspects are performed using line profiles. The corresponding data is presented in Tables \ref{FeIdata} and \ref{FeIIdata}.

\begin{table*}
\caption{Atomic data of \FeI\ lines used in the analysis of the reference stars. Column 1 gives the multiplet number, column 2 the wavelength in \AA, column 3 the lower excitation energy, column 4 the differential log\,$gf$ value with respect to log\,$\varepsilon$(Fe)$_\odot$\,=\,7.51, column 5 the employed log\,C$_6$ value. The products log\,$gf\varepsilon_\odot$ that the log\,$gf$ values presented here imply differ slightly from those presented in Paper II. This is because a global minimization was performed in Paper II to determine the log\,C$_6$ values, whereas lines were fitted individually here on the basis of the minimization procedure.
A radial-tangential approximation to macroturbulence (Gray \cite{Gray77}) was used for the analysis of the Sun, a Gaussian otherwise. The equivalent widths (in m\AA) in columns 6\,--\,10 are of low quality, but only enter the analysis in the derivation of the microturbulence. We discourage from using these values for abundance analyses}
\label{FeIdata}
\begin{center}
{\small
\begin{tabular}{rccrcrrrrr}
mult. & $\lambda$ & $E_{\rm low}$ & log\,$gf_{\rm diff}$ & log\,C$_6$ & Procyon & Gmb~1830 & HD~19445 & HD~84937 & HD~140283\\[3mm]
\hline\\[1mm]
   1 & 5166.2 & 0.000 & $-$4.18 & $-$32.22 &  73  &      &  16  &   5  &   11 \\
   1 & 5225.5 & 0.110 & $-$4.74 & $-$32.20 &  33  &  62  &      &      &      \\
   1 & 5247.0 & 0.087 & $-$4.91 & $-$32.21 &  27  &  57  &      &      &      \\
   1 & 5250.2 & 0.121 & $-$4.88 & $-$32.20 &  26  &  57  &      &      &      \\
   2 & 4347.2 & 0.000 & $-$5.47 & $-$32.16 &      &  27  &      &      &      \\
   2 & 4427.3 & 0.052 & $-$2.94 & $-$32.15 &      &      &  60  &  36  &   56 \\
   2 & 4445.4 & 0.087 & $-$5.43 & $-$32.15 &  11  &  27  &      &      &      \\
   3 & 4232.7 & 0.110 & $-$4.96 & $-$32.12 &      &  47  &      &      &      \\
  13 & 6498.9 & 0.958 & $-$4.64 & $-$32.08 &      &  27  &      &      &      \\
  15 & 5269.5 & 0.859 & $-$1.34 & $-$32.04 &      &      & 111  &  84  &   97 \\
  15 & 5328.0 & 0.915 & $-$1.49 & $-$32.03 &      &      & 100  &  74  &   89 \\
  15 & 5371.4 & 0.958 & $-$1.70 & $-$32.02 &      &      &  89  &  64  &   80 \\
  15 & 5397.1 & 0.915 & $-$2.03 & $-$32.03 &      &      &  73  &  49  &   64 \\
  15 & 5405.7 & 0.990 & $-$1.84 & $-$32.02 &      &      &  76  &  51  &   69 \\
  34 & 6581.2 & 1.485 & $-$4.73 & $-$31.96 &      &   7  &      &      &      \\
  34 & 6739.5 & 1.557 & $-$4.92 & $-$31.91 &      &   4  &      &      &      \\
  36 & 5194.9 & 1.557 & $-$2.15 & $-$31.83 & 109  &      &  34  &  17  &      \\
  36 & 5216.2 & 1.608 & $-$2.22 & $-$31.82 & 101  &      &  27  &  13  &   22 \\
  41 & 4404.7 & 1.557 & $-$0.18 & $-$31.71 &      &      &      &  98  &      \\
  42 & 4147.6 & 1.485 & $-$2.15 & $-$31.67 & 100  &      &      &  17  &   26 \\
  42 & 4271.7 & 1.485 & $-$0.25 & $-$31.70 &      &      &      & 103  &      \\
  62 & 6151.6 & 2.176 & $-$3.31 & $-$31.72 &  20  &  22  &      &      &      \\
  62 & 6297.8 & 2.223 & $-$2.70 & $-$31.72 &  46  &  49  &      &      &      \\
  64 & 6082.7 & 2.223 & $-$3.60 & $-$31.70 &      &  13  &      &      &      \\
  64 & 6240.6 & 2.223 & $-$3.30 & $-$31.71 &  21  &  21  &      &      &      \\
  66 & 5079.2 & 2.198 & $-$2.01 & $-$31.58 &  92  &  93  &      &      &      \\
  66 & 5198.7 & 2.223 & $-$2.14 & $-$31.59 &  77  &      &      &      &      \\
  66 & 5250.6 & 2.198 & $-$2.00 & $-$31.61 &  90  &  87  &      &      &      \\
  68 & 4494.5 & 2.198 & $-$1.17 & $-$31.45 &      &      &      &      &   36 \\
  69 & 4442.8 & 2.176 & $-$2.79 & $-$31.44 &      &  37  &      &      &      \\
  71 & 4282.4 & 2.176 & $-$1.06 & $-$31.39 &      &      &  57  &  41  &   45 \\
 111 & 6421.3 & 2.279 & $-$1.98 & $-$31.71 &  87  &  91  &      &      &      \\
 111 & 6663.4 & 2.424 & $-$2.42 & $-$31.67 &  60  &  54  &      &      &      \\
 111 & 6750.1 & 2.424 & $-$2.56 & $-$31.68 &  45  &  46  &      &      &      \\
 111 & 6978.8 & 2.484 & $-$2.44 & $-$31.67 &  53  &      &      &      &      \\
 114 & 4924.7 & 2.279 & $-$2.15 & $-$31.52 &  73  &  69  &      &      &      \\
 114 & 5049.8 & 2.279 & $-$1.31 & $-$31.54 &      &      &  33  &  18  &   25 \\
 114 & 5141.7 & 2.424 & $-$2.22 & $-$31.50 &  62  &      &      &      &      \\
 116 & 4439.8 & 2.279 & $-$3.04 & $-$31.39 &  24  &  24  &      &      &      \\
 152 & 4187.0 & 2.449 & $-$0.66 & $-$30.79 &      &      &  62  &  43  &   54 \\
 152 & 4222.2 & 2.449 & $-$0.98 & $-$30.80 &      &      &  42  &  26  &   34 \\
 152 & 4233.6 & 2.482 & $-$0.70 & $-$30.79 &      &      &  59  &  40  &   44 \\
 152 & 4250.1 & 2.469 & $-$0.49 & $-$30.81 &      &      &  72  &  48  &   53 \\
 152 & 4260.4 & 2.399 & $-$0.01 & $-$30.84 &      &      & 100  &  74  &   75 \\
 168 & 6393.6 & 2.433 & $-$1.46 & $-$31.65 & 101  &      &  22  &  13  &   15 \\
 168 & 6494.9 & 2.404 & $-$1.25 & $-$31.67 &      &      &  38  &  20  &   28 \\
 168 & 6593.8 & 2.433 & $-$2.32 & $-$31.66 &  54  &  57  &      &      &      \\
 169 & 6136.6 & 2.453 & $-$1.39 & $-$31.62 &      &      &  26  &  15  &   18 \\
 169 & 6191.5 & 2.433 & $-$1.46 & $-$31.63 &      &      &  25  &  13  &   18 \\
 169 & 6252.5 & 2.404 & $-$1.62 & $-$31.64 &  95  &      &  18  &  11  &   12 \\
 170 & 5916.2 & 2.453 & $-$2.90 & $-$31.59 &  25  &      &      &      &      \\
\end{tabular}
}
\end{center}
\end{table*}

\addtocounter{table}{-1}

\begin{table*}
\begin{center}
{\small
\begin{tabular}{rccrcrrrrr}
\multicolumn{5}{l}{\ \ \,continuation of Table \ref{FeIdata}} & & & & &\\[2mm]
mult. & $\lambda$ & $E_{\rm low}$ & log\,$gf_{\rm diff}$ & log\,C$_6$ & Procyon & Gmb~1830 & HD~19445 & HD~84937 & HD~140283\\[3mm]
\hline\\[1mm]
 206 & 6609.1 & 2.559 & $-$2.62 & $-$31.62 &  34  &  33  &      &      &      \\
 207 & 6065.4 & 2.608 & $-$1.46 & $-$31.55 &  97  &  99  &  16  &   7  &   11 \\
 207 & 6137.6 & 2.588 & $-$1.31 & $-$31.57 &      &      &      &      &   15 \\
 207 & 6200.3 & 2.608 & $-$2.35 & $-$31.57 &  44  &  41  &      &      &      \\
 207 & 6230.7 & 2.559 & $-$1.16 & $-$31.59 &      &      &  30  &  15  &   21 \\
 207 & 6322.6 & 2.588 & $-$2.35 & $-$31.58 &  48  &  44  &      &      &      \\
 209 & 5778.4 & 2.588 & $-$3.51 & $-$31.53 &   7  &      &      &      &      \\
 268 & 6546.2 & 2.758 & $-$1.59 & $-$31.54 &  81  &  88  &      &      &    8 \\
 268 & 6592.9 & 2.727 & $-$1.47 & $-$31.55 &      &  92  &      &      &    9 \\
 268 & 6677.9 & 2.692 & $-$1.33 & $-$31.57 & 100  &      &  20  &  11  &   12 \\
 318 & 4890.7 & 2.875 & $-$0.44 & $-$30.80 &      &      &  50  &  33  &   38 \\
 318 & 4891.4 & 2.851 & $-$0.17 & $-$30.81 &      &      &  67  &  48  &   54 \\
 318 & 4918.9 & 2.865 & $-$0.38 & $-$30.81 &      &      &  57  &  36  &   41 \\
 318 & 4920.5 & 2.832 &  0.06   & $-$30.83 &      &      &  81  &  58  &   64 \\
 318 & 4957.2 & 2.851 & $-$0.42 & $-$30.84 &      &      &  56  &  34  &   39 \\
 318 & 4957.5 & 2.808 &  0.19   & $-$30.85 &      &      &  95  &  67  &   75 \\
 342 & 6229.2 & 2.845 & $-$2.94 & $-$31.47 &  19  &  10  &      &      &      \\
 342 & 6270.2 & 2.858 & $-$2.57 & $-$31.47 &  28  &  20  &      &      &      \\
 383 & 5068.7 & 2.940 & $-$1.11 & $-$30.82 &      &      &  18  &  11  &   14 \\
 383 & 5139.4 & 2.940 & $-$0.59 & $-$30.83 &      &      &  44  &  27  &   21 \\
 383 & 5191.4 & 3.038 & $-$0.61 & $-$30.80 &      &      &  36  &  19  &   25 \\
 383 & 5232.9 & 2.940 & $-$0.13 & $-$30.86 &      &      &  67  &  45  &   50 \\
 383 & 5266.5 & 2.998 & $-$0.43 & $-$30.83 &      &      &  47  &  29  &   34 \\
 383 & 5281.7 & 3.038 & $-$0.91 & $-$30.82 &      &      &  22  &  13  &   14 \\
 384 & 4787.8 & 2.998 & $-$2.62 & $-$30.70 &  18  &      &      &      &      \\
 553 & 5217.3 & 3.211 & $-$1.03 & $-$30.71 &  88  &  95  &  10  &   6  &    7 \\
 553 & 5253.4 & 3.283 & $-$1.57 & $-$31.31 &  56  &  44  &      &      &      \\
 553 & 5324.1 & 3.211 & $-$0.05 & $-$30.75 &      &      &  50  &  32  &   36 \\
 553 & 5339.9 & 3.266 & $-$0.62 & $-$30.71 &      &      &  21  &  13  &   13 \\
 553 & 5393.1 & 3.241 & $-$0.65 & $-$30.75 &      &      &  20  &  11  &   13 \\
 554 & 4736.7 & 3.211 & $-$0.80 & $-$30.48 & 110  &      &  23  &  13  &   14 \\
 686 & 5586.7 & 3.368 & $-$0.07 & $-$30.71 &      &      &  42  &  27  &   28 \\
 686 & 5569.6 & 3.417 & $-$0.54 & $-$30.66 &      &      &  21  &  12  &   13 \\
 686 & 5572.8 & 3.396 & $-$0.28 & $-$30.68 &      &      &  32  &  19  &   21 \\
 686 & 5615.6 & 3.332 &  0.05   & $-$30.74 &      &      &  52  &  33  &   37 \\
 686 & 5624.5 & 3.417 & $-$0.75 & $-$30.68 & 100  &      &  20  &   8  &    9 \\
 816 & 6232.6 & 3.654 & $-$1.21 & $-$30.67 &  61  &  44  &      &      &      \\
 816 & 6246.3 & 3.602 & $-$0.71 & $-$30.71 &  89  &  93  &      &      &    6 \\
 816 & 6400.0 & 3.602 & $-$0.25 & $-$30.74 &      &      &  25  &  15  &   17 \\
 816 & 6411.6 & 3.654 & $-$0.62 & $-$30.71 &  98  &      &  13  &   8  &    9 \\
 984 & 4985.2 & 3.928 & $-$0.71 & $-$30.08 &  86  &  74  &      &      &      \\
1031 & 5491.8 & 4.186 & $-$2.22 & $-$30.53 &   6  &      &      &      &      \\
1062 & 5525.5 & 4.230 & $-$1.28 & $-$30.05 &  36  &      &      &      &      \\
1087 & 5662.5 & 4.178 & $-$0.60 & $-$30.07 &  75  &  56  &      &      &      \\
1087 & 5705.4 & 4.301 & $-$1.49 & $-$30.05 &  21  &   9  &      &      &      \\
1092 & 5133.6 & 4.178 &  0.20   & $-$30.40 &      &      &  25  &  15  &   16 \\
1094 & 5074.7 & 4.220 & $-$0.10 & $-$30.32 & 104  &  89  &  14  &   9  &      \\
1146 & 5364.8 & 4.445 &  0.15   & $-$30.21 & 108  &  94  &  16  &  11  &    9 \\
1146 & 5367.4 & 4.415 &  0.22   & $-$30.25 &      &      &  21  &  14  &   11 \\
1146 & 5369.9 & 4.371 &  0.34   & $-$30.31 &      &      &  25  &  17  &   15 \\
1146 & 5383.3 & 4.312 &  0.42   & $-$30.38 &      &      &  30  &  21  &   20 \\
1146 & 5424.0 & 4.320 &  0.52   & $-$30.39 &      &      &  34  &  22  &   22 \\
1164 & 5410.9 & 4.473 &  0.14   & $-$30.20 &      & 102  &  19  &  11  &   13 \\
1164 & 5415.1 & 4.386 &  0.41   & $-$30.31 &      &      &  29  &  18  &   17 \\
1178 & 6024.0 & 4.548 &  0.03   & $-$30.40 &  93  &  71  &  10  &   6  &    5 \\
1179 & 5855.1 & 4.607 & $-$1.56 & $-$30.26 &  12  &      &      &      &      \\
1195 & 6752.7 & 4.638 & $-$1.27 & $-$30.05 &  20  &      &      &      &      \\[1mm]
\hline
\end{tabular}
}
\end{center}
\end{table*}

\addtocounter{table}{1}

\begin{table*}
\caption{Atomic data of \FeII\ lines used in the analysis of the reference stars. Column 1 gives the multiplet number, column 2 the wavelength in \AA, column 3 the lower excitation energy, column 4 the differential $gf$ value with respect to log\,$\varepsilon$(Fe)$_\odot$\,=\,7.51, column 5 the employed log\,C$_6$ value. A radial-tangential approximation to macroturbulence (Gray \cite{Gray77}) was used for the analysis of the Sun, a Gaussian otherwise. The equivalent widths (in m\AA) in columns 6\,--\,10 are of low quality, but only enter the analysis in the derivation of the microturbulence. We discourage from using these values for abundance analyses}
\label{FeIIdata}
\begin{center}
{\small
\begin{tabular}{rccrcrrrrr}
mult. & $\lambda$ & $E_{\rm low}$ & log\,$gf_{\rm diff}$ & log\,C$_6$ & Procyon & Gmb~1830 & HD~19445 & HD~84937 & HD~140283\\[3mm]
\hline\\[1mm]
27 & 4416.8 & 2.77 & $-$2.64 & $-$31.78 &     &  28 &  14 &  14 &  11  \\
27 & 4233.1 & 2.57 & $-$2.04 & $-$31.78 &     &  77 &  44 &  47 &  47  \\
35 & 5136.8 & 2.83 & $-$4.44 & $-$31.78 &  20 &     &     &     &      \\
35 & 5132.6 & 2.79 & $-$4.17 & $-$31.78 &  35 &     &     &     &      \\
35 & 5100.6 & 2.79 & $-$4.18 & $-$31.78 &  37 &     &     &     &      \\
36 & 4993.3 & 2.79 & $-$3.73 & $-$32.18 &  49 &     &     &     &      \\
37 & 4629.3 & 2.79 & $-$2.38 & $-$32.18 &     &     &  20 &  19 &  17  \\
37 & 4582.8 & 2.83 & $-$3.20 & $-$32.18 &  86 &     &     &     &      \\
37 & 4555.8 & 2.82 & $-$2.48 & $-$32.18 &     &  33 &  18 &  19 &  17  \\
37 & 4515.3 & 2.83 & $-$2.53 & $-$31.88 &     &  32 &  14 &  15 &  14  \\
37 & 4491.4 & 2.84 & $-$2.75 & $-$31.78 & 108 &  21 &   9 &   9 &      \\
38 & 4620.5 & 2.82 & $-$3.33 & $-$31.78 &  78 &   8 &     &     &      \\
38 & 4576.3 & 2.83 & $-$2.95 & $-$31.78 &  96 &     &   6 &   6 &      \\
38 & 4508.2 & 2.84 & $-$2.39 & $-$31.78 &     &     &  18 &  20 &  16  \\
40 & 6516.0 & 2.88 & $-$3.29 & $-$32.11 &  79 &     &     &     &      \\
40 & 6432.6 & 2.88 & $-$3.61 & $-$32.11 &  67 &     &     &     &      \\
40 & 6369.4 & 2.88 & $-$4.19 & $-$32.11 &  33 &     &     &     &      \\
41 & 5284.1 & 2.88 & $-$3.11 & $-$32.11 &  88 &  11 &     &     &      \\
42 & 5169.0 & 2.88 & $-$1.28 & $-$32.01 &     &     &  83 &  80 &  70  \\
42 & 5018.4 & 2.88 & $-$1.29 & $-$32.11 &     &  96 &  70 &  72 &  70  \\
42 & 4923.9 & 2.88 & $-$1.53 & $-$31.91 &     &  87 &  61 &  65 &  58  \\
43 & 4656.9 & 2.88 & $-$3.62 & $-$32.11 &  58 &     &     &     &      \\
46 & 6084.1 & 3.19 & $-$3.84 & $-$32.19 &  36 &     &     &     &      \\
46 & 5991.3 & 3.14 & $-$3.56 & $-$32.19 &  51 &     &     &     &      \\
48 & 5362.8 & 3.19 & $-$2.57 & $-$32.19 &     &     &   8 &   8 &   7  \\
48 & 5264.8 & 3.22 & $-$3.08 & $-$32.19 &  73 &   6 &     &     &      \\
49 & 5425.2 & 3.19 & $-$3.27 & $-$32.19 &  65 &     &     &     &      \\
49 & 5325.5 & 3.21 & $-$3.21 & $-$32.19 &  68 &     &     &     &      \\
49 & 5316.6 & 3.14 & $-$1.91 & $-$31.89 &     &  44 &  27 &  29 &  26  \\
49 & 5234.6 & 3.21 & $-$2.21 & $-$31.89 &     &  28 &  15 &  16 &  14  \\
49 & 5197.5 & 3.22 & $-$2.27 & $-$31.89 &     &  23 &  13 &  13 &  11  \\
74 & 6456.3 & 3.89 & $-$2.09 & $-$32.18 & 100 &  11 &   6 &   7 &      \\
74 & 6416.9 & 3.87 & $-$2.67 & $-$32.18 &  63 &     &     &     &      \\
74 & 6247.5 & 3.87 & $-$2.33 & $-$32.18 &  89 &   6 &     &     &      \\
74 & 6239.9 & 3.87 & $-$3.49 & $-$32.18 &  29 &     &     &     &      \\
74 & 6149.2 & 3.87 & $-$2.76 & $-$32.18 &  64 &     &     &     &      \\[1mm]
\hline
\end{tabular}
}
\end{center}
\end{table*}

\subsection{Evidence for non-vanishing hydrogen collisions: the subgiant HD~140283}
The dynamic range of \FeI\ abundances in HD~140283 upon varying \SH\ is exemplified by Figure~\ref{depcoeffs140283} and Table~\ref{sh0sh3lte}: the extreme underpopulation of \FeI\ in its atmosphere is clearly seen. The depopulation with respect to the LTE case already starts in deep photospheric layers. For vanishing \SH\ the departure coefficients $b_i$ drop to 0.5 near $\log$\,$\tau$\,=\,$-$0.5.

\begin{figure}
\begin{center}
\includegraphics[angle=90,width=0.47\textwidth,clip=]{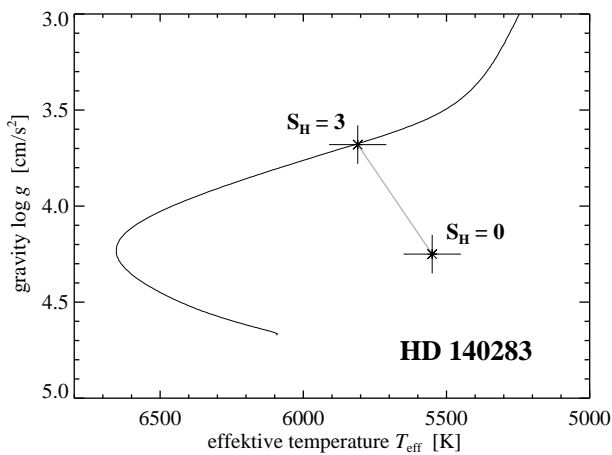}
\caption{A comparison between the stellar parameters of HD~140283 for different assumptions concerning the efficiency of hydrogen collisions. \SH\,=\,3 corresponds to our best estimate and places this star in a subgiant phase of evolution. A mass close to 0.8\,M$_\odot$ is indicated from a track of appropriate metallicity by VandenBerg et al. (\cite{VSRIA00}). However, assuming \SH\,=\,0 (no collisions) results in a \Teff-\logg\ combination in irreconcilable conflict with the star's known properties, both in terms of its evolutionary stage and mass}
\label{newpos140283}
\end{center}
\end{figure}

Models with \SH\,$>$\,0 produce less extreme departures as radiative processes (photoionization) are increasingly compensated for by collisional ones (hydrogen collisions). More importantly, though, a non-vanishing \SH\ secures {\em relative thermal excitation equilibrium}: although strongly overionized with respect to \FeII, the \FeI\ populations are kept thermal relative to one another, even by collisional efficiencies lower than indicated by the Drawin formula (e.g.\ \SH\,=\,0.1). Since only the ratio of the departure coefficients enters the line source function S$^l_\nu$, identical $b_i$ result in a Planckian S$^l_\nu$. This means that within a given model non-LTE corrections  only depend on the contribution function, i.e. the depth of formation. Thus low-excitation lines will experience larger corrections on average.

Calculations including hydrogen collisions then constitute a model sequence with non-LTE abundance corrections steadily decreasing with an increasing value for \SH. The behaviour found in the metal-poor stars is morphologically similar to the case of the Sun. We thus refer the reader to Papers I \& II for more details.

Table~\ref{sh0sh3lte} shows how these departure coefficients translate into \FeI\ abundances. With a 1$\sigma$ scatter of 0.05\,dex in \FeII\ the \SH\,=\,0 model is ruled out at more than 3$\sigma$. In addition to the discrepant mean values, the  \SH\,=\,0 model would require a significantly smaller microturbulence for \FeI\ ($\xi_{\rm FeI}$\,=\,1.2\,km/s), in conflict with the one derived from \FeII\ lines ($\xi_{\rm FeII}$\,=\,1.7\,km/s). If the stellar parameters were determined assuming \SH\,=\,0 (this time using H$\alpha$ as the temperature indicator which shows some dependence on gravity), the ionization equilibrium would be reached at \Teff\,=\,5550\,K, \logg\,=\,4.25 and [Fe/H]\,=\,$-$2.26 (cf. Fig. \ref{newpos140283}). Assuming a sensible stellar mass of $M$\,=\,0.8\,M$_\odot$, \DHIP\ (as defined in the caption of Table \ref{stellarparameters}) would be $-$54\,\%.

\subsection{The other metal-poor reference stars}
On the Balmer profile temperature scale Gmb~1830, HD~19445 and HD~84937 all require an \SH\ close to 3. This is surprising as these stars are in different evolutionary phases and require different non-LTE corrections. In fact, the non-LTE effects in Gmb~1830 are smaller than in the Sun resulting in an abundance and gravity correction with an opposite sign from all the other stars. This turns out to reduce the star-to-star scatter significantly (cf. Fig.~\ref{mycalib}). Non-LTE effects for HD~19445 are as large as in the Sun. The differential non-LTE result thus coincides with the LTE one.

With \SH\ set to 3 the non-LTE corrections are generally quite small and never entail a gravity correction larger than 0.1\,dex. It might come as a surprise that we are nonetheless able to remove the large discrepancies between Mg\,{\sc i}b and \FeI/\FeII\ as gravity indicators uncovered by Fuhrmann (\cite{Fuhrmann98a}). He points out that using \FeI/\FeII\ results in gravities up to 0.5\,dex smaller than those from Mg\,{\sc i}b which are in excellent agreement with \HIP. The solution to this apparent contradiction lies in the choice of C$_6$ values: The larger C$_6$ values as proposed by Anstee \& O'Mara (\cite{AOM91}, \cite{AOM95}) enter our analysis via the evaluation of the strong solar \FeI\ lines where a certain exchange between {\em f} values and damping constants is possible (for more details on our exact choice of damping parameters see Paper II). Thus our set of {\em differential f} values differs systematically from that employed in the studies by Fuhrmann (\cite{Fuhrmann98a}) and Fuhrmann (\cite{Fuhrmann98}, \cite{Fuhrmann03}) which employ C$_6$ values calculated using the Uns\"old ({\cite{Unsold68}) approximation.

We have no physical explanation for our empirical finding that hydrogen collisions are as efficient as \SH\,=\,3. Laboratory measurements for the \ion{Na}{i} resonance line at H beam energies above 10\,\eV\ indicate efficiencies as low as \SH\,$\approx$\,0.01 (Fleck et al. \cite{FGS91}). We are convinced that \SH\,=\,3 is not an artefact of our modelling, as the same code results in much lower efficiencies for simpler atoms like Mg or Al (Baum\"uller \& Gehren \cite{BG96}, Zhao et al. \cite{ZBG98}). But whereas \MgI\ and \AlI\ are essentially shaped by a few strong photoionization cross-sections, nearly all \FeI\ terms are depopulated by photoionization. This gives \FeI\ a unique atomic structure. In spite of the laboratory results mentioned above, it can thus not be ruled out that hydrogen collisions among \FeI\ terms are in fact as efficient as the Drawin (\cite{Drawin68}, \cite{Drawin69}) formula predicts (or three times as efficient). We note that \SH\ would even be higher, if we hadn't enforced upper-level thermalization above 7.3\,\eV\ to account for term incompleteness (for details see Paper I).

Holweger (\cite{Holweger96}) uses the red giant Pollux ($\equiv \gamma$ Gem) to calibrate \SH\ to a value of 0.1. This result is not in direct conflict with ours as Holweger did not have access to the Bautista (\cite{Bautista97}) cross-sections for photoionization.

\begin{table}[!t]
\caption[]{Different sets of stellar parameters for Procyon. The radius is determined from \Teff\ and $M_{\rm bol}$ and has to be compared with the fundamental value of $R$\,=\,(2.07 $\pm$ 0.02)\,R$_\odot$. $\Delta_{\rm \,HIP}\,=\unboldmath\,100\,(d_{\rm\,spec}\,-\,d_{\rm \,HIP})/d_{\rm \,HIP}$}
\label{Procyon}
\begin{center}
\begin{tabular}{ccccccr}
model & \Teff\ & $\log g$ & [\FeII/H]$_{\rm diff}$ & mass & radius & $\Delta_{\rm HIP}$ \\
 & [K] & & & [M$_\odot$] & [R$_\odot$] & [\%] \\
\noalign{\smallskip}\hline \noalign{\smallskip}
1 & 6512 & 3.96 & \m0.09 & 1.42 & 2.13 &\m3.1\\
2 & 6508 & 3.96 & \m0.03 & 1.42 & 2.13 &\m3.0 \\
3 & 6508 & 3.81 & \m0.10 & 1.42 & 2.13 &+15.0 \\
4 & 6600 & 3.96 & \m0.03 & 1.42 & 2.06 &\m0.1 \\
\noalign{\smallskip}\hline \noalign{\smallskip}\\
\end{tabular}
\end{center}
\vspace*{-0.5cm}
\hspace*{0.52cm} 1 \ \Teff, \logg, [Fe/H]: Allende Prieto et al. (\cite{AAGL02})\\
\hspace*{0.52cm} 2 \ \Teff: Balmer profiles/\logg: astrometry/[Fe/H]: from \FeII\\
\hspace*{0.52cm} 3 \ \Teff: Balmer profiles/\logg: \FeI/\FeII\,(\SH\,=\,3)\\
\hspace*{0.52cm} 4 \ \Teff: free parameter, \logg: astrometry {\em and\/} \FeI/\FeII\,(\SH\,=\,3)\\
\end{table}

From the point of view of hydrodynamical model atmospheres it could be argued that the model temperatures in the outer layers of static atmospheres are overestimated (Asplund et al. \cite{ANTS99}). Lower temperatures would predominantly make low-excitation lines of neutral species come out stronger, thus alleviating the need for strongly thermalizing collisions (high values of \SH) to counterbalance photoionization. Using 2D radiation-hydrodynamical simulations, Steffen \& Holweger \cite{SH02} derive granulation corrections for lines arising from low-lying levels of neutral species reaching up to $-$0.3\,dex in the solar case. More work on combining hydrodynamical with non-LTE computations is clearly needed (cf. Shchukina \& Trujillo Bueno (\cite{STB01}) for first attempts with respect to Fe).

\subsection{Procyon}\label{procyon}
Apart from the Sun, $\alpha$ CMi is one of the very few stars for which the whole set of fundamental stellar parameters except metallicity is known from direct measurements. Allende Prieto et al. (\cite{AAGL02}) review recent determinations of these quantities. Its proximity ($d$\,=\,(3.5 $\pm$ 0.01)\,pc) and the presence of a white dwarf companion (Procyon B) allow to constrain Procyon's mass ($M$\,=\,(1.42 $\pm$ 0.06)\,M$_\odot$). From the radius ($R$\,=\,(2.07 $\pm$ 0.02)\,R$_\odot$) and measurements of the integrated flux the effective temperature is known to within $\sim$\,100\,K and probably lies between 6500\,K and 6600\,K. Mass and radius can be combined to yield the logarithmic surface gravity, \logg\,=\,3.96 $\pm$ 0.02. The metallicity is close to solar. Allende Prieto et al.'s final choice for their comparative 1D static and 3D hydrodynamical modelling is \Teff\,=\,6512\,K, \logg\,=\,3.96 and solar composition. We adopt the same value for the surface gravity and derive the effective temperature from the Balmer lines which yield an average value of \Teff\,=\,(6508 $\pm$ 60)\,K. The evolutionary tracks of VandenBerg et al. (\cite{VSRIA00}) point towards a somewhat higher mass of 1.52\,M$_\odot$, in better agreement with the mass estimate of Girard et al. (\cite{GWL00}) who also consider ground-based measurements for the angular separation of Procyon A and B. To be able to compare our results directly to those of Allende Prieto et al. (\cite{AAGL02}), we stick to the choice of parameters given above. We note, however, that this choice of mass produces a 3\,\% offset with respect to \HIP.

\begin{table*}
\caption{Stellar parameters derived using Fe\,{\sc i/ii} in non-LTE as a gravity indicator with S$_{\rm H}$\,=\,3. The spectroscopic distance $d_{\rm spec}$ is calculated from $\log\pi_{\rm spec}=0.5\,([g]-[M])-2\,[T_{\rm eff}]-0.2\,(V+{\rm BC}+A_{V}+0.25)$, where [$X$]\,=\,$\log (X/X_\odot)$. $\Delta_{\rm\,HIP}\,=\unboldmath\,100\,(d_{\rm\,spec}\,-\,d_{\rm \,HIP})/d_{\rm \,HIP}$. The oxygen abundance is derived by means of profile analysis of the IR triplet in non-LTE (Reetz \cite{Reetz99}), the magnesium abundance from profiles of weak optical lines in non-LTE ($\lambda\lambda$ 4571, 4702, 4730, 5528, 5711\,\AA, Zhao et al. \cite{ZBG98})} \label{stellarparameters}
\begin{center}
\begin{tabular}{rrrrrrrrrrrr}  object     &  $T_{\rm eff}$ &$\!\log\,g\!$&     [Fe/H] & [O/Fe] & [Mg/Fe] & mass & $A_V$ & ${\rm BC}$ & $d_{\rm\,spec}$ & $d_{\rm\,HIP}$ & $\Delta_{\rm HIP}$ \\[1mm]      & [K] &  &$\!$ NLTE & NLTE & NLTE &[M$_\odot$] & [mag] & [mag] &[pc] & [pc] & [\%] \\[1mm]
\noalign{\smallskip}\hline \noalign{\smallskip}\\[-3mm]
HD 103095 &5070& 4.66 & $-$1.36  & $+0.63$ & $+0.21$ & 0.62 &  0.00 &  $-$0.32 &    8.86 &    9.16 & $-$3.2 \\[1mm]
HD19445 &6032& 4.40 & $-$2.08  & $+0.68$ & $+0.49$ & 0.67 &  0.00 &  $-$0.21 &   38.80 &   38.68 &  0.3 \\[1mm]
HD 140283 &5806& 3.68 & $-$2.43  & $+0.71$ & $+0.26$ & 0.79 &  0.13 &  $-$0.23 &   56.60 &   57.34 & $-$1.3 \\[1mm]
HD 84937 &6346& 4.00 & $-$2.16  & $+0.59$ & $+0.39$ & 0.79 &  0.11 &  $-$0.18 &   80.34 &   80.39 & $-$0.1\\[1mm]
\noalign{\smallskip}\hline \noalign{\smallskip}
\end{tabular}
\end{center}
\end{table*}

\begin{figure}
\begin{center}
\vspace*{.5cm}
\includegraphics[angle=90,width=0.47\textwidth,clip=]{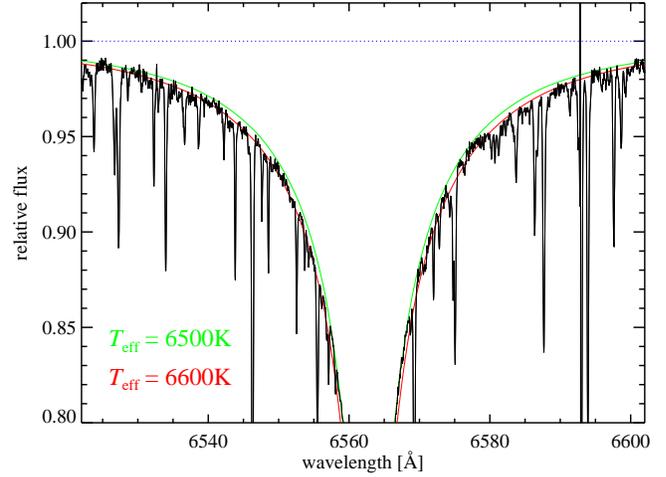}
\caption{Profile fits to H$\alpha$ in the FOCES spectrum of Procyon. The revised temperature (\Teff\,=\,6600\,K) is in disagreement with the observations}
\label{ProcyonHalpha}
\end{center}
\end{figure}

It turns out that our model is not able to fulfil the ionization equilibrium of iron with \SH\,=\,3. With differential non-LTE effects in abundance of merely 0.01\,dex the unbalance amounts to $\Delta\,($\FeII$\,-\,$\FeI$)$\,=\,0.07\,dex. To remove this discrepancy this model would require a gravity of \logg\,=\,3.81, many standard deviations away from the astrometric result.\\
How would one have to tune \SH\ to remove $\Delta\,($\FeII$\,-\,$\FeI$)$? Surprisingly, no choice of \SH\ is able to fulfil the ionization equilibrium. Using the largest non-LTE corrections possible (\SH\,=\,0) the discrepancy is merely reduced by 0.02\,dex. In this sense Procyon only documents our failure to use this star as a calibrator for \SH.

Seemingly, Allende Prieto et al.~(\cite{AAGL02}) are more successful with their 1D LTE modelling of Procyon's iron spectrum. Their $\Delta\,($\FeII$\,-\,$\FeI$)$ amounts to 0.02\,dex in $\log\,\varepsilon$(Fe) ($\log\,\varepsilon$(\FeI)\,=\,7.30 $\pm$ 0.11 vs. $\log\,\varepsilon$(\FeII)\,=\,7.32 $\pm$ 0.08) or 0.04\,dex in \logg. Upon closer inspection one notices that these values are {\em non-differential} abundances. Formulated relative to the Sun where a discrepancy of 0.06\,dex was obtained by these authors, the discrepancy in Procyon rises to 0.08\,dex which is in complete numerical agreement with our LTE result. Irrespective of whether one prefers differential analyses or not, we conclude that Allende Prieto et al.~(\cite{AAGL02}) do not master iron in this star, either (their 3D hydrodynamical modelling results in a residual trend of $\log\,\varepsilon$(\FeI) with line strength which could be removed by including a microturbulence of $\xi$\,$\approx$\,0.3\,km/s).

Steffen (\cite{Steffen85}) investigated different ionization equilibria in Procyon and drew the conclusion that the fundamental temperature needs to be revised upwards by more than 200\,K. Alternatively, the gravity would have to be lowered to \logg\,$\simeq$\,3.55 which can be ruled out by astrometry. In the framework of  non-LTE modelling, a similar temperature correction would be required to bring [O\,{\sc i} 6300] into concordance with the IR triplet lines (Reetz \cite{Reetz99}). We can follow along this track and seek the temperature at which the \SH\,=\,3 model brings $\log$\,$\varepsilon$(\FeI) into concordance with $\log$\,$\varepsilon$(\FeII) without contradicting the astrometric distance constraint. This temperature is found to be \Teff\,=\,6600\,K or 90\,K higher than our best spectroscopic estimate based on Balmer profiles. As can be appreciated from inspecting Fig.~\ref{ProcyonHalpha} this temperature is in conflict with the Balmer profile analysis. It is, however, in not in disagreement with the gravity inferred from the Mg\,{\sc i}b lines when employing the non-LTE model atom for magnesium presented by Zhao et al. (\cite{ZBG98}). This is because the strong-line method is not very sensitive to changes in temperature. Thus, while disquieting on the whole, our result is an improvement over the analysis of Steffen (\cite{Steffen85}). Higher temperatures for Procyon certainly seem possible in view of the Str\"{o}mgren photometry of Edvardsson et al. (\cite{EAGLNT93}, \Teff\,=\,6705\,K) and the IRFM temperature of Alonso et al. (\cite{AAMR96}, \Teff\,=\,6631\,K). From the new self-broadening theory for Balmer lines presented by Barklem et al. (\cite{BPO00}) higher temperatures (\Teff\,=\,6570\,K) are possible, {\em if\/} the zero-point offset for the Sun is properly accounted for.

\section{Stellar parameters from the S\boldmath$_{\rm H}$\unboldmath\,=\,3 model}\label{sh3}
Having calibrated \SH, we reanalyse the iron spectra of the reference stars with our non-LTE model for iron. The final para\-meters are given in Table~\ref{stellarparameters}.

Fig.~\ref{mycalib} compares the stellar parameters in non-LTE with their LTE counterparts. In particular, the scatter is reduced in going from LTE to non-LTE, as differential non-LTE effects for HD~103095 bear the opposite sign to those of all the other stars. Non-LTE corrections are largest for HD~84397 and HD~140283. A positive slope of a linear regression line in this diagram would thus indicate underestimated non-LTE correction and vice versa. This effect will be seen in the analyses of Th\'evenin \& Idiart (\cite{TI99}) and Fulbright (\cite{Fulbright00}) below.

\begin{figure}[!t]
\begin{center}
\vspace*{.5cm}
\includegraphics[angle=0,width=0.47\textwidth,clip=]{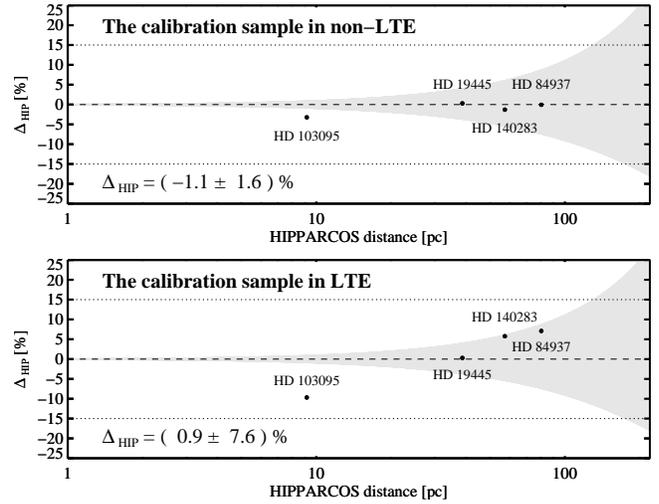}
\caption{\DHIP\ as a function of the astrometric distance, both in non-LTE (top) and LTE (bottom). The grey wedge indicates the typical uncertainty of the \HIP\ parallaxes. When non-LTE gravities are considered the star-to-star scatter is significantly reduced }
\label{mycalib}
\end{center}
\end{figure}


\subsection{Comparison with other studies}
Based on our newly established methodology, we can now compare the resulting spectroscopic distance scale with that of other studies. We compare the residual offset with respect to the \HIP\ astrometry and the star-to-star scatter for the --~admittedly~-- small sample of reference stars. For a fair and valid comparison, we use the published stellar parameters, but rederive bolometric corrections and masses. Since an estimate for the [$\alpha$/Fe] ratio is needed for interpolating among appropriate evolutionary tracks, we only consider studies which supply an $\alpha$-element abundance alongside [Fe/H]. We therefore disregard e.g. the study of Soubiran et al. (\cite{SKC98}).

Particular attention is paid to studies which also undertook kinetic equilibrium calculations for iron. As can be appreciated from comparing the upper panel of Fig.~\ref{mycalib} to Fig.~\ref{distcomp}, all studies using \FeI/\FeII\ are an order of magnitude less precise than our calibration.

\subsubsection{Gratton et al. (\cite{GCEG99})}
These authors employ a rather simplistic model atom of Takeda (\cite{Takeda91}), treat the \FeI\ photoionization and background opacities empirically and calibrate \SH\ by means of RR Lyrae stars. As \SH\ is determined to be 30, thermal populations result for their stars which is why the LTE stellar parameters of Gratton et al. (\cite{GCC96}) are still the basis for the analysis of Carretta et al. (\cite{CGS00}). Due to the dominant influence of oxygen on the morphology of the evolutionary tracks, we use their published [O/Fe]$_{\rm NLTE}$ as [$\alpha$/Fe].

The topmost panel of Fig.~\ref{distcomp} shows the resulting spectroscopic distance scale in comparison with the astrometric distances of \HIP. Since the temperatures these authors derive via the IRFM are in excellent agreement with ours, it is surprising to see that the gravities are somewhat overestimated rather than underestimated. From the work of Fuhrmann et al. (\cite{FPFRG97}) we would have expected the LTE gravities to be significantly too low (cf. Introduction). Since we have shown this effect to be mainly due to an inappropriate choice of damping parameters C$_6$, we assume that a similar effect is at work in the analyses of Gratton et al. (\cite{GCC96}). More worrying than the 12\,\% offset is the large scatter, e.g. in the multiple analysis of HD~140283 which amounts to nearly 30\,\%.

\subsubsection{Th\'evenin \& Idiart (\cite{TI99})}
In an attempt to model the kinetic equilibrium of iron {\em ab initio}, Th\'evenin \& Idiart (\cite{TI99}) put together a comprehensive and up-to-date model atom which is quite comparable to the one used by us. Photoionization is treated according to the calculations of Bautista (\cite{Bautista97}), UV line opacities are crudely represented by a scaling rule for continuous opacities. In combination with their neglect of hydrogen collisions (\SH\,$\equiv$\,0), these assumptions lead to very large non-LTE corrections. For example, the gravity correction for HD~140283 is calculated to be 0.54\,dex. We use the larger value of [Mg/Fe]$_{\rm NLTE}$ and [Ca/Fe]$_{\rm NLTE}$ from Idiart \& Th\'evenin (\cite{IT00}) as an estimate of [$\alpha$/Fe].

On the temperature scale employed by these authors, the non-LTE corrections are clearly overestimated, severely so in the case of HD~84937 (\DHIP\,=\,\m34\,\%). While the scatter is comparable to that of Gratton et al. (\cite{GCEG99}), the offset is significantly larger: the sample distances are incompatible with \HIP\ at the 2$\sigma$ level.

These authors also analyse Procyon. At a temperature of \Teff\,=\,6631\,K the ionization equilibrium of iron is reached at a gravity of \logg\,=\,4.00 without departures from LTE. Judging from our own calculations (cf. Sect.~\ref{procyon}) this modelling success is not surprising and basically the result of choosing a high effective temperature. The derived parameters yield \DHIP\,=\,\m3.3\,\%. Including this star into the sample, the mean offset would be reduced to 17\,\% with the scatter increasing to 12\,\%.

\begin{figure}
\begin{center}
\vspace*{.5cm}
\includegraphics[angle=0,width=0.47\textwidth,clip=]{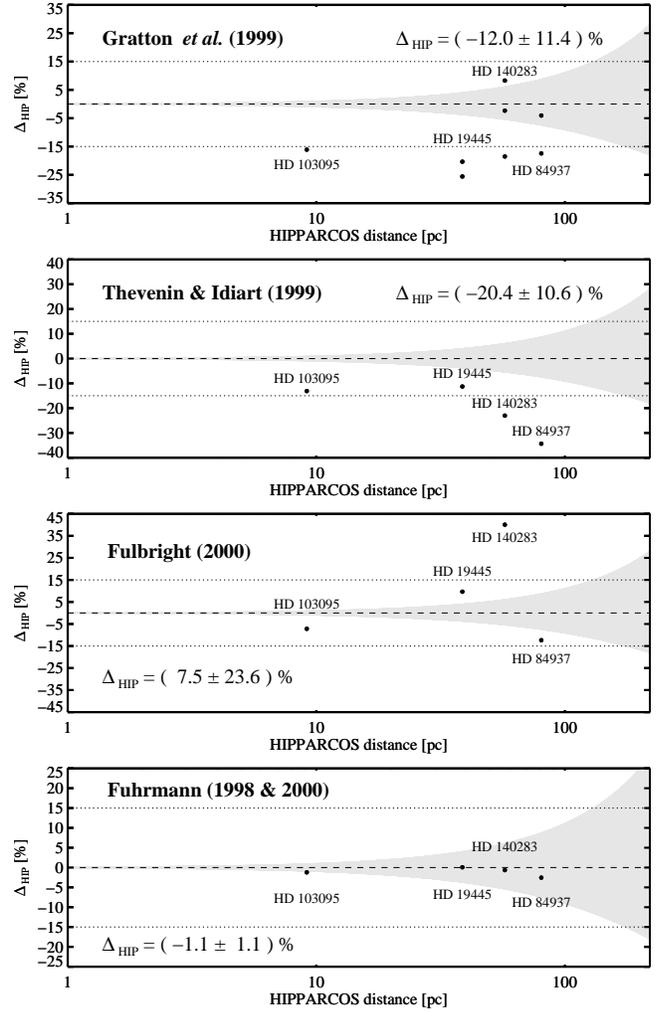}
\caption{Same as Fig.~\ref{mycalib}, here for the studies discussed in the text. Note the different y-axis scales. Except for the work of Fuhrmann (\cite{Fuhrmann98}, \cite{Fuhrmann00}), all studies show large star-to-star scatter and/or offsets with respect to \HIP}
\label{distcomp}
\end{center}
\end{figure}

\subsubsection{Fulbright (\cite{Fulbright00})}
This analysis of 168 mostly metal--poor stars rests almost entirely on iron as a plasmadiagnostic tool: the temperature determination is based on the LTE excitation equilibrium of \FeI, gravity and metallicity are inferred from the LTE ionization equilibrium of \FeI/\FeII. These steps are iterated until $\log\,\varepsilon$(\FeI) matches $\log\,\varepsilon$(\FeII) to "within 0.03 dex in most cases". At best, we can therefore expect an accuracy for \logg\ of 0.05\,dex. We use [Mg/Fe]$_{\rm LTE}$ as a proxy for [$\alpha$/Fe]. The masses we derive from Fulbright's stellar parameters show a wide range: HD~103095 would be a very low-mass object ($M$\,=\,0.45\,M$_\odot$), HD~140283 a star of nearly solar mass ($M$\,=\,0.95\,M$_\odot$). This might indicate unrealistic combinations of \Teff\ and \logg.

While the sample mean comes out relatively close to \HIP, the star-to-star scatter is very large. The offset for HD~140283 is 40\,\%. Statistically speaking, the gravities are only know to within 0.25\,dex. Just like in the case of Th\'evenin \& Idiart (\cite{TI99}) not a single star falls in the 1$\sigma$ uncertainty range of the \HIP\ astrometry. All this means that Fulbright's methods fail to give consistent results for metal-poor F and G stars.

\subsubsection{Fuhrmann (\cite{Fuhrmann98}, \cite{Fuhrmann00})}
These studies are methodologically closest to our own, in the sense that Fuhrmann employs the same model atmosphere and the same Balmer profile temperature scale. The pressure-broadened wings of Mg\,{\sc i}b in LTE are used as a gravity indicator with the remarkable result of \DHIP\,=\,(2\,$\pm$\,5)\,\% for 100 stars. Balmer profiles and Mg\,{\sc i}b lines are practically orthogonal methods to determine \Teff\ and \logg\ which helps to prevent disadvantageous propagation of errors. Masses are interpolated among tracks of Bernkopf (\cite{Bernkopf98}). The stellar parameters of HD~140283 are revised with respect to Fuhrmann (\cite{Fuhrmann98a}) and are private communication.

When using the tracks of VandenBerg et al. (\cite{VSRIA00}) the resulting distances are slightly shorter than with the Bernkopf (\cite{Bernkopf98}) tracks. The internal consistency is, however, still very impressive. Fuhrmann's work clearly demonstrates the capabilities of a careful, differential analysis in LTE.

We note in passing that the revision of the stellar mass for Procyon (down to $M$\,=\,1.42\,M$_\odot$) worsens the spectroscopic result published by Fuhrmann (\cite{Fuhrmann98}) to \DHIP\,=\,\m9.6\,\%. Thus Procyon still is a paramount test case for any spectroscopic analysis.

\section{Extreme applications: HE~0107--5240 \& CS~29497--004}
In the meantime, our model has been applied to a couple of new extreme halo objects: HE~0107\m5240, the most iron-deficient star currently known (Christlieb et al. \cite{Christlieb02}) and CS~29497\m004, a new highly $r$-process enhanced star (Christlieb et al. \cite{Christlieb03}). Since both objects are on the upper red giant branch (RGB), non-LTE corrections to gravity are larger than in the case of the reference stars discussed above.\\
Both stars have effective temperatures around 5100\,K and surface gravities around 2.2. Despite their vastly different metallicities ([Fe/H]\,=\,\m5.3 vs. \m2.6), non-LTE corrections to gravity are very similar and amount to +0.3\,dex. They are similar because two competing processes cancel each other. On the one hand, the \FeI\ departure coefficients of HE~0107\m5240 are more extreme than those of CS~29497\m004; on the other, the \FeI\ lines in HE~0107\m5240 are much weaker and therefore originate deeper in the atmosphere. As test calculations have shown, hydrogen collisions are still important to consider at these RGB gravities.

Further calculations indicate that non-LTE effects in \FeI\ reach a plateau value below a metallicity comparable to that of the most metal-poor globular clusters found in our Galaxy. Globular cluster giants are thus ideal targets to verify the predictions of our calculations. It is encouraging to see that at intermediate metallicities these predictions are in good agreement with the mild \FeI/\FeII\ imbalance found in the LTE ana\-lyses of giants in M5 and M71 by Ram\'{i}rez \& Cohen (\cite{RC03}).

\section{Conclusions}
We have presented a procedure to determine the poorly-known efficiency of collisions with neutral hydrogen in our kinetic equilibrium calculations for the formation of \FeI\ and \FeII\ lines in the atmospheres of solar-type stars. This procedure rests on exploiting the astrometric constraint of \HIP\ parallaxes. The scaling factor \SH\ is determined to be 3 which makes hydrogen collisions the most important atomic process counterbalancing the otherwise overwhelming photoionization in metal-poor stars. We emphasize, however, that our goal has not been to determine the absolute strength of hydrogen collisions. Rather, our analysis is aimed at removing biases that are present in the analysis of metal-poor stars when using the iron ionization equilibrium in LTE using 1D model atmospheres.

For a representative sample of local halo stars the combination of Balmer profile temperatures and non-LTE iron ionization equilibrium gravities succeeds in fulfilling the astrometric constraint. Other authors (Gratton et al. \cite{GCEG99}, Th\'evenin \& Idiart \cite{TI99}, Fulbright \cite{Fulbright00}) present results an order of magnitude less accurate. This underlines our claim that hydrogen collisions must be considered in kinetic equilibrium calculations. The independent method to derive \logg\ used by Fuhrmann (\cite{Fuhrmann98}, \cite{Fuhrmann00}) yields comparable results.

Our method cannot establish the ionization equilibrium of Procyon at the fundamental parameters \Teff\,=\,6510\,K and \logg\,=\,3.96. An upward revision of Procyon's temperature by 90\,K (1.3\,\%) to \Teff\,=\,6600\,K is indicated instead, in agreement with the radius determination based on its bolometric magnitude. This temperature is, however, in disagreement with the spectroscopic constraint of the Balmer profiles.

Using our carefully calibrated model of \FeI/\FeII, cool stars over the whole range of metallicities encountered in the Galaxy can now be analysed in a homogeneous way with reduced methodological biases without resorting to trigonometric parallaxes.

\begin{acknowledgements}
AJK wishes to thanks the {\sl Studienstiftung des deutschen Volkes} for continuous financial support between 1999\,--\,2002. Illuminating discussions with Paul Barklem, Jan Bernkopf, Klaus Fuhrmann, Frank Grupp, Lyudmila Mashonkina and Johannes Reetz are gratefully acknowledged. The referee is thanked for several helpful suggestions which have been incorporated in the manuscript.
\end{acknowledgements}

\end{document}